\documentstyle[aps]{revtex}
\input epsf

\def\jepsfbox#1{\typeout{#1} \epsfbox{#1}}

\def\plottwo#1#2{\centering \leavevmode
\epsfxsize=.45\columnwidth \jepsfbox{#1} \hfil
\epsfxsize=.45\columnwidth \jepsfbox{#2}}
\def\plotthree#1#2#3{\centering \leavevmode
\epsfxsize=.30\columnwidth \jepsfbox{#1} \hfil
\epsfxsize=.30\columnwidth \jepsfbox{#2} \hfil
\epsfxsize=.30\columnwidth \jepsfbox{#3}}
\def\jcite#1#2{#1 \cite{#2}}
\def\tilde{\mathaccent"365}			% tilde accent
\def\eg{{\it e.g.~}}
\def\ie{{\it i.e.~}}
\def\cf{{\it c.f.~}}

\def\rmmat#1{{\hbox{\rm #1}}}
\def\rmscr#1{\rmmat{\scriptsize #1}}
\newcommand{\be}{\begin{equation}}
\newcommand{\ee}{\end{equation}}
\newcommand{\ba}{\begin{eqnarray}}
\newcommand{\ea}{\end{eqnarray}}
\def\figref#1{Fig.~\ref{fig:#1}}
\def\eqref#1{Eq.~\ref{eq:#1}}
\begin{document}
%\draft
%
\newcommand{\bfi}{{\bf B}} \newcommand{\efi}{{\bf E}}
\newcommand{\kfi}{{\bf k}}
\newcommand{\afi}{{\bf A}}
\newcommand{\lag}{{\cal L}}
\newcommand{\dLIII}{{\frac{\partial^3\lag}{\partial I^3}}}
\newcommand{\dLII}{{\frac{\partial^2\lag}{\partial I^2}}}
\newcommand{\dLI}{{\frac{\partial \lag}{\partial I}}}
\newcommand{\dLKKK}{{\frac{\partial^3 \lag}{\partial K^3}}}
\newcommand{\dLKK}{{\frac{\partial^2 \lag}{\partial K^2}}}
\newcommand{\dLK}{{\frac{\partial \lag}{\partial K}}}
\newcommand{\dLIK}{{\frac{\partial^2 \lag}{\partial I \partial K}}}
\newcommand{\dLJJ}{{\frac{\partial^2 \lag}{\partial J^2}}}
\newcommand{\dLIJ}{{\frac{\partial^2 \lag}{\partial I \partial J}}}
\newcommand{\dLJ}{{\frac{\partial \lag}{\partial J}}}
\newcommand{\dLJJJ}{{\frac{\partial^3 \lag}{\partial J^3}}}
\newcommand{\dLIJJ}{{\frac{\partial^3 \lag}{\partial I \partial J^2}}}
\newcommand{\dLIIJ}{{\frac{\partial^3 \lag}{\partial I^2 \partial J}}}
\title{Electron-Positron Jets from a Critically Magnetized Black Hole}
\author{Jeremy S. Heyl\footnote{Chandra Postdoctoral Fellow}}
\address{Theoretical
Astrophysics 130-33, California Institute of Technology,
Pasadena, California 91125 \\
{\rm and}\\
Harvard Observatory, MS-51, 
60 Garden Street, Cambridge 02138}

\maketitle
\begin{abstract}
The curved spacetime surrounding a rotating black hole dramatically
alters the structure of nearby electromagnetic fields.  The Wald field
which is an asymptotically uniform magnetic field aligned with the angular
momentum of the hole provides a convenient starting point to analyze
the effects of radiative corrections on electrodynamics in curved
spacetime.  Since the curvature of the spacetime is small on the
scale of the electron's Compton wavelength, the tools of quantum field
theory in flat spacetime are reliable and show that a rotating black
hole immersed in a magnetic field approaching the quantum critical
value of $B_k=m^2 c^3/(e\hbar) \approx 4.4 \times 10^{13}$~G $\approx
1.3\times10^{-11}$~cm$^{-1}$ is unstable.  Specifically, a maximally rotating 
three-solar-mass black hole immersed in a magnetic field of $2.3
\times 10^{12}$~G would be a copious producer of electron-positron
pairs with a luminosity of $3 \times 10^{52}$ erg s$^{-1}$.
\end{abstract}

\section{Introduction}

The recent discovery that gamma-ray bursts (GRBs) are associated with
galaxies at cosmological distances has spurred the development of
thoeretical models of the central engines of these objects which emit
$\sim 10^{51-53}$~ergs (assuming isotropic emission) over the span of
several to several hundred seconds.  Some of the more popular models
involve some sort of electromagnetic bomb: a quickly rotating,
strongly magnetized neutron star (\eg
\cite{1992Natur.357..472U,1998ApJ...498L..31B}) or a rotating black
hole threaded by a strong magnetic field (\eg
\cite{1997ApJ...482L..29M,1998ApJ...494L..45P,Lee00}).  The first
model extends the standard picture of radio pulsar spindown (\eg
\cite{Gold69}) to ultrastrong magnetic fields and high spin
frequencies; the second model recasts the Blandford and Znajek
\cite{Blan77} mechanism for the central engine of quasars in the realm
of a stellar black hole accreting the debris of a tidally disrupted
neutron star.

An examination of the instability of the magnetized vacuum surrounding
a rotating black hole provides an excellent starting point to
understanding these processes.  Van Putten has studied the analogue to
Hawking radiation for a rotating, magnetized black hole and finds that
if the applied field approaches the quantum electrodynamic critical
value of $4.4\times 10^{13}$~G \cite{vanP00a,vanP00b}, a rotating
stellar-mass black hole will produce $\sim 10^{49}$ erg s$^{-1}$ in
pairs.  Although this technique based on Hawking radiation provides an
estimate of the pair production near the hole, the pair
production for a strongly magnetized, stellar mass black hole depends
only extremely weakly on the Hawking temperature of the hole or
equivalently on the spacetime curvature near the horizon; therefore,
accurate results may be obtained if one ignores the effects
of spacetime curvature on the quantum mechanics of the electromagnetic
field surrounding the hole.  Specifically, Gibbons \cite{Gibb75}
argues that if the mass, $M$, of black hole greatly exceeds
$10^{17}$~g, the quantum mechanical effects of spacetime curvature may
safely be ignored for particles more massive than an electron.

Gibbons \cite{1976MNRAS.177P..37G} examined the problem of how an
uncharged rotating black hole embedded in a magnetic field will
acquire a charge ($Q=2BJ$, the Wald charge\cite{Wald74}) through
pair-creation near the horizon.  The current paper builds on Gibbons's
picture \cite{1976MNRAS.177P..37G} and examines in detail the
pair-creation process after the hole has acquired the Wald charge.

The spacetime curvature does affect the structure of the applied
electromagnetic field on scales comparable to that of the black
hole.  This paper begins with a treatment of this effect in
\S~\ref{sec:wald} through a discussion of the the Wald\cite{Wald74}
field for spacetimes which admit both timelike and axial Killing
vectors.  \S~\ref{sec:kerr} specializes this discussion to the
spacetime surrounding a rotating black hole, the Kerr spacetime
\cite{Wald74,Land2,Hawk73}. \S~\ref{sec:efflagr} calculates the pair
production rate in a locally inertial frame threaded by both an electric
and magnetic field based on the Heisenberg-Euler lagrangian
\cite{Heis36,Itzy80,Heyl97hesplit}.  This theoretical basis is
utilized to calculate both numerically (\S~\ref{sec:numer}) and
analytically (\S~\ref{sec:analy}), the pair production near a
rotating, magnetized black hole.

\section{Theoretical Background}

\subsection{The Wald Field}
\label{sec:wald}

In a vacuum spacetime, a linear combination of Killing vectors yields
a solution for the electromagnetic vector potential also in vacua.  The
spacetime surrounding a rotating blackhole yields two Killing vectors
$\psi$ and $\eta$, corresponding to rotations about the angular
momentum axis and time translations.  Wald \cite{Wald74} found that if
the electromagnetic field asymptotically becomes a uniform magnetic
field, the vector potential is given by
\be
A_\mu = \frac{1}{2}B_0 \left (\psi_\mu + \frac{2 J}{M} \eta_\mu \right
) - \frac{Q}{2 M} \eta_\mu
\ee
where $Q$ is the charge of the hole.  The electromagnetic field is
assumed to be a test field, \ie it does not curve spacetime.  If the
scalar potential of the horizon differs from that at infinity, the
hole preferentially accretes charge until the potentials are equal.
\newcommand{\QW}{Q_\rmscr{W}}
This occurs for $\QW=2 B_0 J$ and
\be
A_\mu = \frac{1}{2} B_0 \psi_\mu.
\ee
During the production of the electron-positron jets, the charge of the
hole may depart from the Wald value; therefore, if $Q'=Q-\QW$, the
complete vector potential is given by
\be
A_\mu = \frac{1}{2} B_0 \psi_\mu - \frac{Q'}{2 M} \eta_\mu.
\ee

\subsection{The Kerr Geometry}
\label{sec:kerr}

Since the calculation of the electromagnetic field near the black
hole focusses on the Killing vectors, it is propitious to use the
Boyer-Lindquist coordinates in which the metric takes the form
\cite{Wald74,Land2,Hawk73},
\be
ds^2 = - \left ( 1 - \frac{2 M r}{\Sigma} \right ) dt^2 
	-  \frac{4 M a r \sin^2 \theta}{\Sigma} dt d\phi
	+ \left [ \frac{(r^2+a^2)^2 - \Delta a^2 \sin^2
\theta}{\Sigma} \right ] \sin^2\theta d\phi^2 
	+ \frac{\Sigma}{\Delta} d r^2 + \Sigma d \theta^2
\ee
where
\ba
\Sigma &=& r^2 + a^2 \cos^2 \theta \\
\Delta &=& r^2 +a^2 - 2 M r
\ea
and the Killing vectors are $\eta^\mu=[1,0,0,0]$ and
$\psi^\mu=[0,0,0,1]$.  In these coordinates, the field tensor 
is simply related to the derivatives of the
metric.  If the charge of the hole differs from the Wald value, the
tensor consists of two components,
\be
F_{\mu\nu} = \frac{B_0}{2} F^{(\psi)}_{\mu\nu} - \frac{Q'}{2 M} F^{(\eta)}_{\mu\nu}
\ee
where
\be
F^{(\psi)}_{\mu\nu} = \left [ 
\begin{array}{cccc}
   0      & g_{03,1} & g_{03,2} &     0     \\
-g_{03,1} &    0     &     0    & -g_{33,1} \\
-g_{03,2} &    0     &     0    & -g_{33,2} \\
   0      & g_{33,1} & g_{33,2} &     0     
\end{array}
\right ]
\ee
and
\be
F^{(\eta)}_{\mu\nu} = \left [ 
\begin{array}{cccc}
   0      & g_{00,1} & g_{00,2} &     0     \\
-g_{00,1} &    0     &     0    & -g_{03,1} \\
-g_{00,2} &    0     &     0    & -g_{03,2} \\
   0      & g_{03,1} & g_{03,2} &     0     
\end{array}
\right ].
\ee
The invariants $I$ and $J$ also depend simply on the metric
coefficients and their derivatives,
\ba
I &=& \frac{1}{-g} \left [ \frac{B_0^2}{2} I^{(\psi\psi)} + \frac{(Q')^2}{2 M^2} I^{(\eta\eta)}
- \frac{B_0 Q'}{M} I^{(\psi\eta)} \right ]\\
J &=& \frac{1}{-g} \left [ B_0^2 J^{(\psi\psi)} + \frac{(Q')^2}{M^2} J^{(\eta\eta)}
- \frac{B_0 Q'}{M} J^{(\psi\eta)} \right ] \\
\ea
where $g=\det(g_{\mu\nu})$ and
\ba
I^{(\psi\psi)} &=& 
-g_{03,1}^2 g_{22} g_{33} + 2 g_{03,1} g_{22} g_{03} g_{33,1} 
-g_{03,2}^2 g_{11} g_{33} + 2 g_{03,2} g_{11} g_{03} g_{33,2}
-g_{33,1}^2 g_{22} g_{00} - g_{33,2}^2 g_{11} g_{00}   \\
I^{(\eta\eta)} &=& 
-g_{00,1}^2 g_{22} g_{33} + 2 g_{00,1} g_{22} g_{03} g_{03,1} 
-g_{00,2}^2 g_{11} g_{33} + 2 g_{00,2} g_{11} g_{03} g_{03,2}
-g_{03,1}^2 g_{22} g_{00} - g_{03,2}^2 g_{11} g_{00}  \\
I^{(\eta\psi)} &=& 
 -g_{03,1}  g_{22} g_{33}  g_{00,1} + g_{03,1}^2  g_{22}  g_{03}
- g_{03,2} g_{11}  g_{33} g_{00,2}  + g_{03,2}^2 g_{11} g_{03}
+ g_{33,1} g_{22} g_{03} g_{00,1}  \\
& & ~~~~~~ 
- g_{33,2} g_{11} g_{00} g_{03,2} 
+ g_{33,2} g_{11} g_{03} g_{00,2}  \\
J^{(\psi\psi)} &=& 
 g_{33,2} g_{03,1} - g_{33,1} g_{03,2} \\
J^{(\eta\eta)} &=& 
 g_{03,2} g_{00,1} - g_{03,1} g_{00,2} \\
J^{(\eta\psi)} &=&   
g_{33,2} g_{00,1} - g_{33,1} g_{00,2} .
\ea

It is important to verify that the various quantities are well behaved
on the horizon.  A useful expression is the value of $I$ at the pole,
\be
I = 2 B_0^2 + 2 \frac{Q'}{M^2} B_0 \sqrt{\frac{2-r_H}{r_H}} +
\frac{1}{2} \left ( \frac{Q'}{M^2} \right )^2 \frac{2-r_H}{r_H}.
\label{eq:poleI}
\ee
$J$ vanishes over the entire horizon and $r_H=M+\sqrt{M^2+a^2}$, the radial 
coordinate of the horizon.

\subsection{The Effective Lagrangian of Quantum Electrodynamics}
\label{sec:efflagr}

For a uniform external field the effective Lagrangian of quantum
electrodynamics may be written as \cite{Heis36,Itzy80,Heyl97hesplit}
\be
\lag = -\frac{1}{4} I + \frac{\alpha}{8\pi^2} B_k^2 \int_0^\infty 
\frac{d\zeta}{\zeta}
e^{-i\zeta} 
\left [  \frac{a b}{B_k^2} \coth\left ( \zeta \frac{a}{B_k} \right) 
\cot \left (\zeta \frac{b}{B_k} \right) -
\frac{1}{\zeta^2} \right ]
\label{eq:lagdef}
\ee
where $-2 (a+ib)^2=I+iJ$, $I=F^{\mu\nu}F_{\mu\nu}$,
$J= {\cal F}^{\mu\nu} F_{\mu\nu}$, 
$B_k=m^2 c^3/(e\hbar) \approx 4.4 \times 10^{13}$~G $\approx 
1.3\times10^{-11}$~cm$^{-1}$ and $m$ is the mass of the electron.
$a$ and $b$ are the strengths of the electric and magnetic field
measured in a reference frame where the two fields are parallel and 
transform as scalars.

The pair production probability ($w$) is simply related to the
imaginary part of the Lagrange density of the electromagnetic field
\cite{Itzy80,Heyl97hesplit}, $w=2(4\pi \hbar)^{-1} \Im \lag$.  The
imaginary part of the integrand in \eqref{lagdef} is even along the
real axis so the range of integration may be extended over the entire
real axis and the contour completed with a semicircle encompassing the
negative imaginary portion of the complex plane.  The integrand has
poles along both the real and imaginary axes; however, since the
integrand is even along the real axis, the residues for the real poles
cancel in pairs leaving \cite{1976PhRvD..14..340D},
\be
w = \frac{1}{\pi} \left(\frac{\hbar}{mc}\right)^{-3} 
\left(\frac{\hbar}{mc^2}\right)^{-1} \frac{a b}{B_k^2} 
\sum_{n=1}^\infty \frac{1}{n} 
\coth\left (\frac{n\pi b}{a}\right) \exp \left (-n \pi \frac{B_k}{|a|}
\right  ).
\label{eq:wrate}
\ee
The more familiar limiting case is where $b \ll a$ which yields \cite{Itzy80,Schw51},
\be
w \approx \frac{1}{\pi^2} \left(\frac{\hbar}{mc}\right)^{-3} 
\left(\frac{\hbar}{mc^2}\right)^{-1} \frac{a^2}{B_k^2} 
\rmmat{dilog} \left [ 1 - \exp(-\pi B_k/|a|) \right ].
\label{eq:wratea}
\ee
Taking the opposite limit yields
\be
w \approx -\frac{1}{\pi} \left(\frac{\hbar}{mc}\right)^{-3} 
\left(\frac{\hbar}{mc^2}\right)^{-1} \frac{a b}{B_k^2} 
\rmmat{log} \left [ 1 - \exp(-\pi B_k/|a|) \right ].
\ee
For a given value of $a$ the pair production rate increases monotonically 
with $b$ and linearly for $b \gg a$.  

\section{Pair-Production near Rotating Black Holes}

If the mass, $M$, of black hole greatly exceeds $m_\rmscr{Planck}^2/m
\approx 10^{17}$~g, the tools of quantum field theory in a flat
spacetime are adequate to describe the pair production rate near the
black hole \cite{Gibb75,1976MNRAS.177P..37G}; combining the results of
the previous sections yields a definitive prediction for the pair
production and emission energy from the vicinity of the black hole
before the magnetosphere forms.

If the magnetic field is parallel (antiparallel) to the angular
momentum, positrons (electrons) tend to escape to infinity, and the
hole quickly acquires a slight negative (positive) charge, so that
equal numbers of each charge escape to infinity.  For a maximally
rotating hole, the bulk of the pair creation occurs between latitudes
of $30^\circ - 50^\circ$.  If the magnetic field is parallel to the
angular momentum of black hole, the positrons escape from the lower
half of that range.  For more slowly spinning holes, the emission
region moves closer to the equator.  The emission rate on the horizon
itself vanishes unless the hole has a significant amount of charge
\cite{Gibb75}, \ie $Q'/M^2 < (-4 + 2\sqrt{3}) B_0$.
\begin{figure}
\plottwo{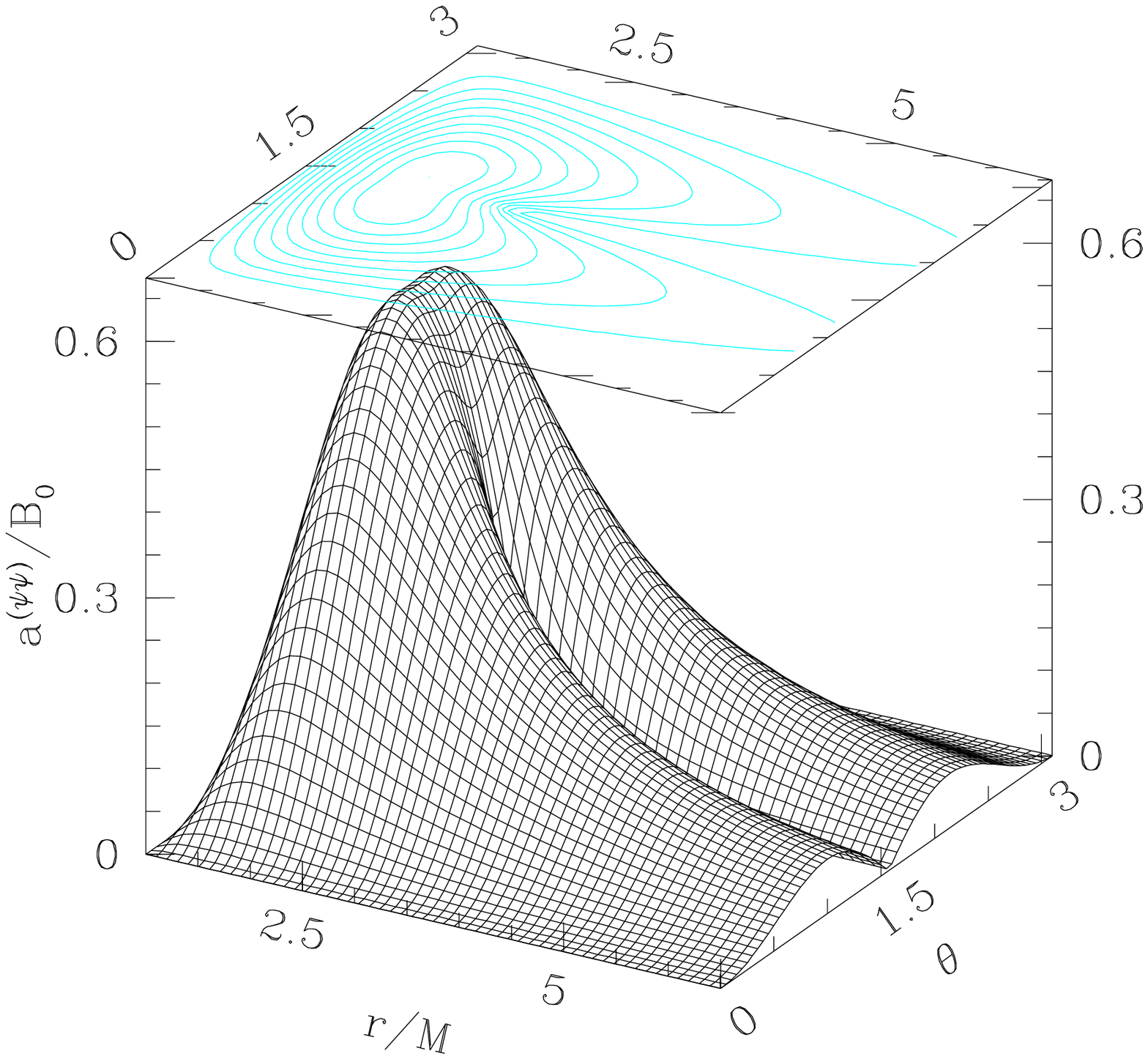}{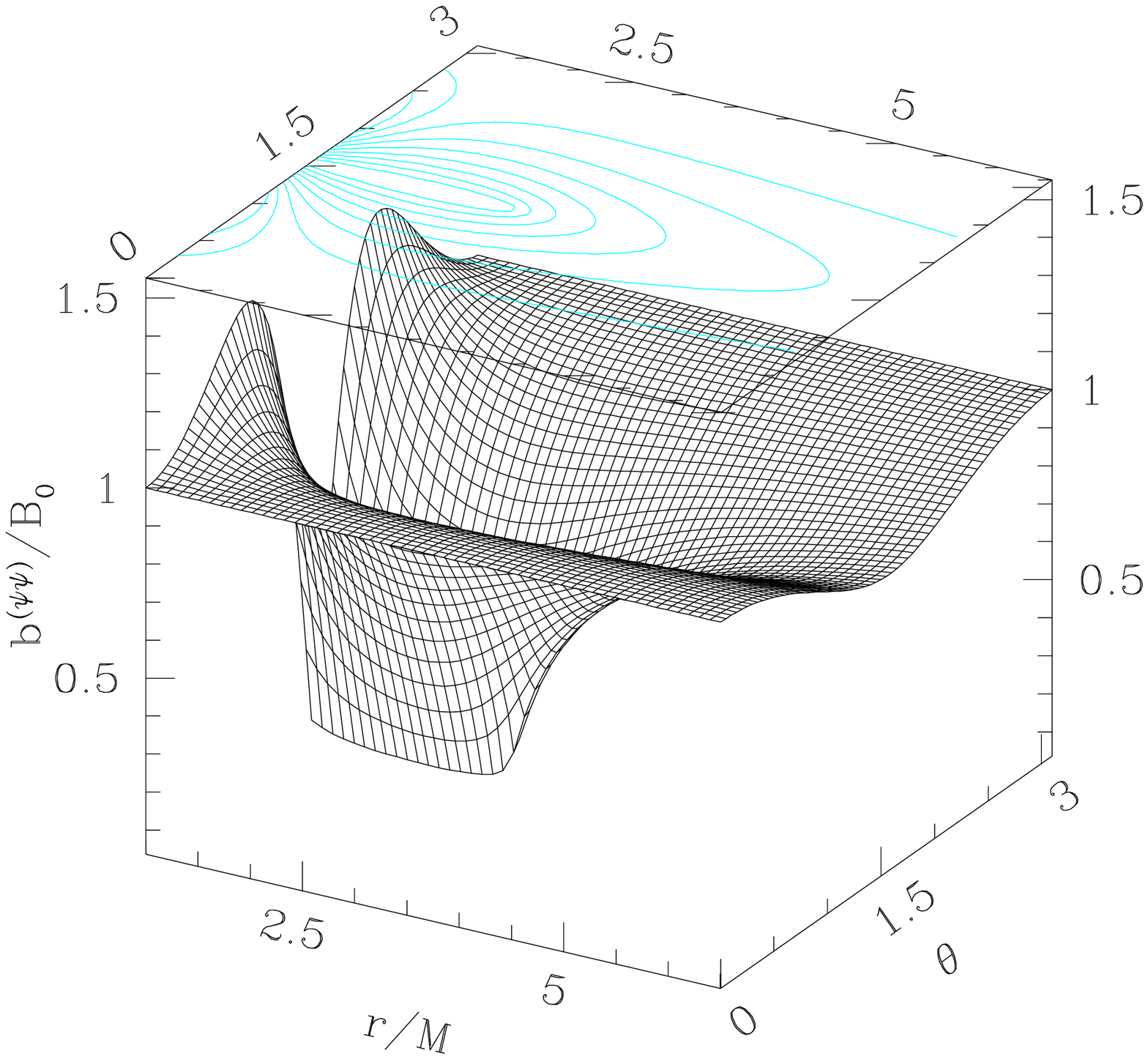}
\caption{The values of the Lorentz invariant ``electric'' and
``magnetic'' components of the electromagnetic field surrounding a
maximally rotating black hole with the Wald charge, $\QW=2 B_0 J$.  }
\label{fig:ap2}
\end{figure}

\figref{ap2} demonstates that the local strength of the electric field
inside the static limit is comparable to that of the applied magnetic field.
In regions where $a \sim B_k$, the pair-production rate is on the
order of $10^{52}$~cm$^{-3}$~s$^{-1}$ or $L \sim 10^2$ (in the rest
mass of the particles alone) over the entire spacetime.  Clearly, pair
production near a rotating black hole must become important for 
$a \ll\ B_k$

\subsection{Numerical Results}
\label{sec:numer}

Summing the pair production rate over a coarse grid yields an an
approximate picture of the pair production near rotating black holes.
Assuming that the particles reach infinity with the electrostatic 
injection energy from the region where they appear, gives an
estimate of the total pair production luminosity.  
\begin{figure}
\plottwo{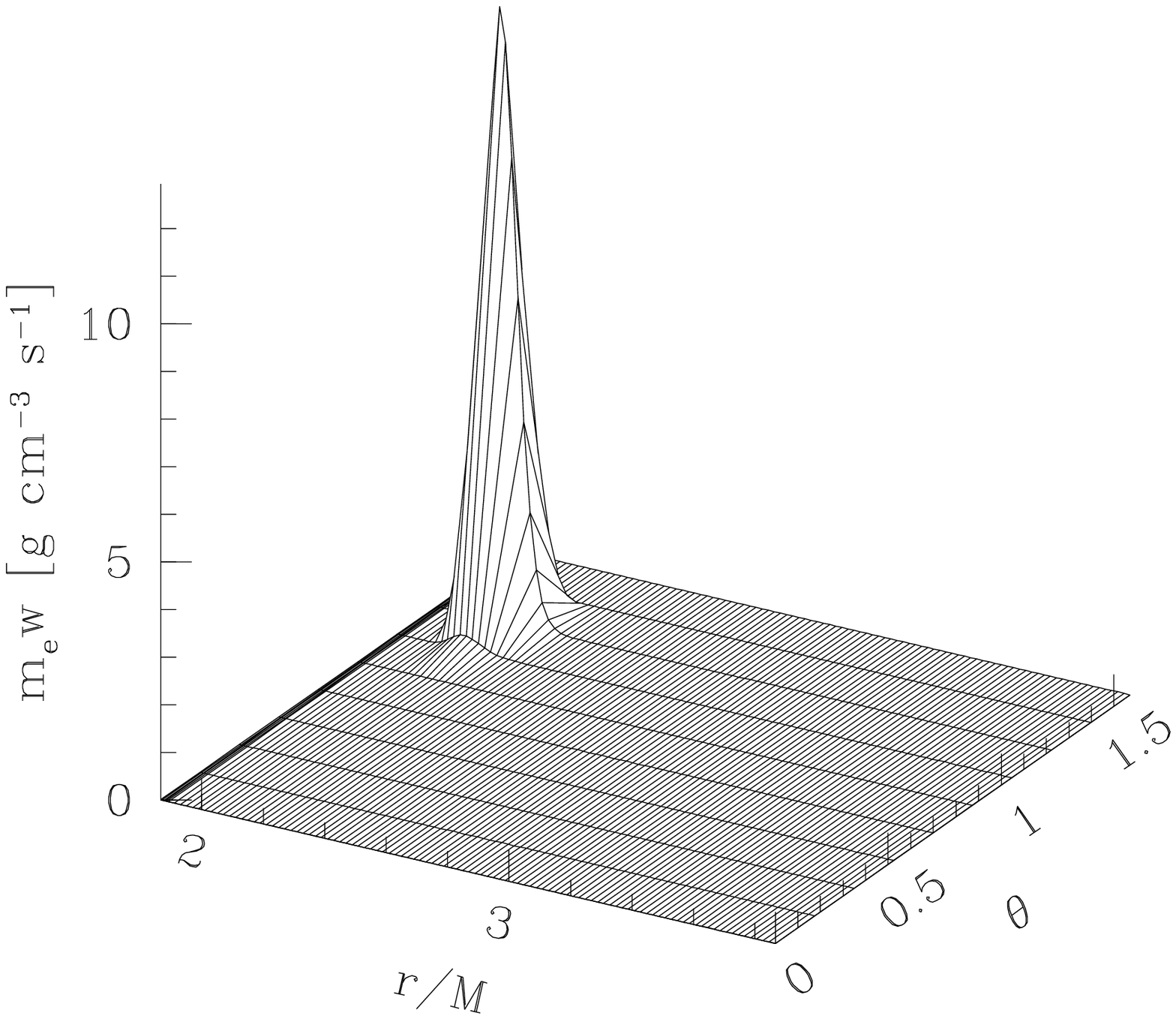}{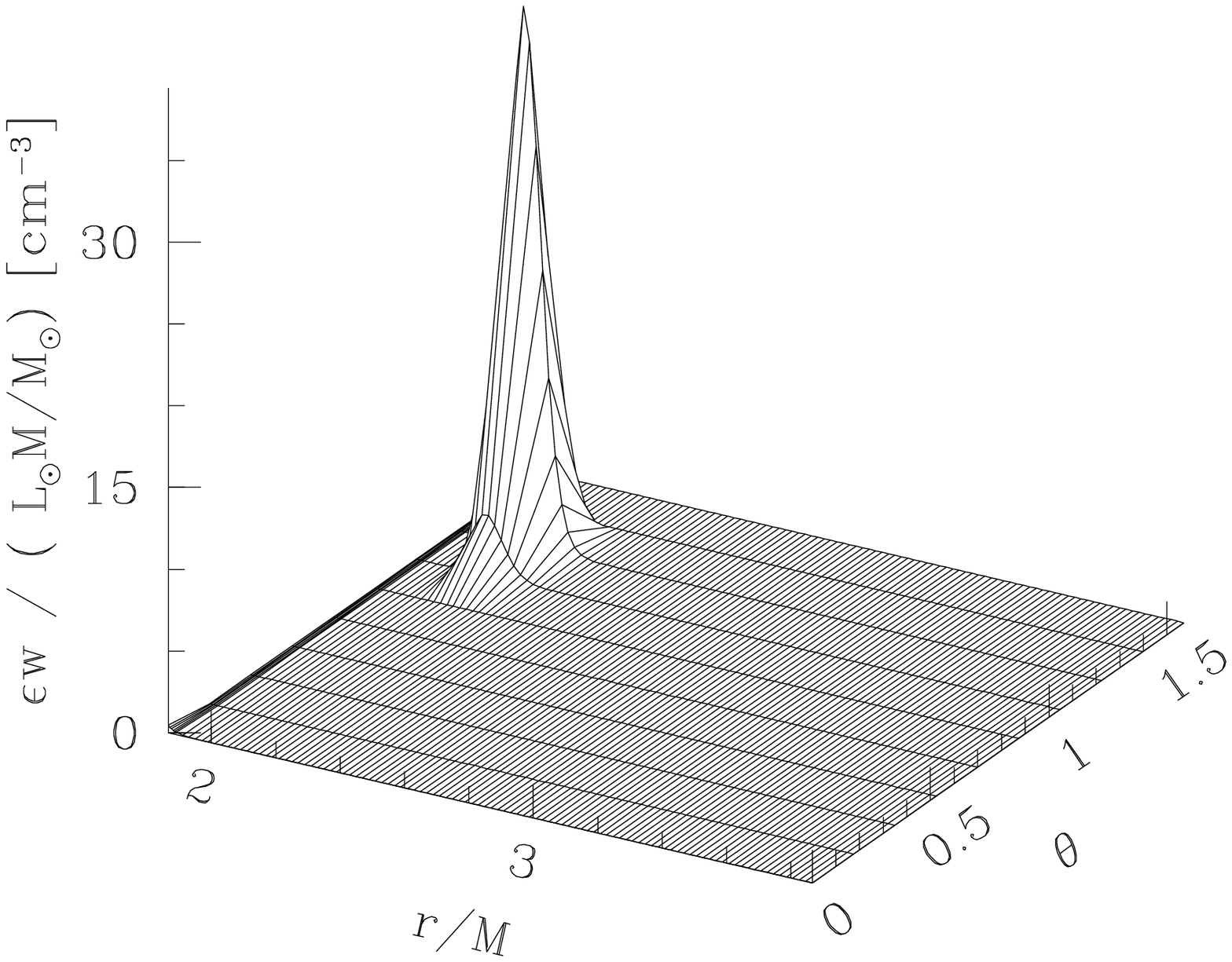}
\caption{The pair-production rate and luminosity for a black hole with
$a = 0.5 M$ and $B_0=0.27 B_k$}
\label{fig:rate5}
\end{figure}
\figref{rate5} depicts the pair-production rate for a black hole with
$a=0.5 M$ with $B_0=0.26 B_k$.  The charge of the black hole deviates
from the Wald value with $Q'/M^2 = -0.2 B_k$ to make the outflow
approximately neutral.  The total charge of the hole is $Q=0.23 \QW$.
The pair production is highly concentrated near $r=2.1 M$ and a
latitude of twenty degrees.  This is significantly above the horizon
which lies at $r\approx 1.86 M$ and slightly outside the static limit.
The luminosity is double peaked since the zero potential surface runs
through the peak of the pair production.

\begin{table}
\caption{Summary of pair-production luminosities from subcritically
rotating black holes (numerical results for $L=10^{50} (M/1 \rmmat{km})^4$ erg s$^{-1}$).}
\label{tab:numermodels}
\begin{tabular}{cc|ccccc}
$a/M$ & $B_0/B_k$ & $Q'/(M^2 B_k)$ & $\QW/(M^2 B_k)$ & $W/|W|$ \\ \hline
0.1 & 0.93 & -0.17 & 0.19 & 0.038 \\ 
0.2 & 0.55 & -0.20 & 0.22 & 0.144 \\ 
0.3 & 0.40 & -0.21 & 0.24 & -0.218 \\ 
0.4 & 0.33 & -0.23 & 0.26 & 0.217 \\ 
0.5 & 0.27 & -0.24 & 0.27 & -0.004 \\ 
0.6 & 0.23 & -0.24 & 0.28 & -0.227 \\ 
0.7 & 0.20 & -0.24 & 0.28 & 0.059 \\ 
0.8 & 0.16 & -0.21 & 0.26 & -0.085 \\ 
0.9 & 0.14 & -0.14 & 0.25 & 0.276 \\ 
1.0 & 0.05 & -0.06 & 0.10 & -0.040 
\end{tabular}
\end{table}

Since the pair production is highly localized, especially for slowly
rotating black holes, the estimates of the total luminosity are rather
sensitive to the resolution of the grid (higher resolutions may yield
higher luminosities), and it is difficult to estimate the value of
$Q/M^2$ required to achieve strict charge neutrality in the outflow.

\subsection{Analytic Treatment}
\label{sec:analy}

Since the pairs are produced in a small region of spacetime around the
black hole where $a \ll\ M$, several important simplifications are
available.  First only the first term in \eqref{wrate} will be
important.  Second, the pair-production rate near the peak is
approximately Gaussian with characteristic widths in the $r-$ and
$\theta-$directions.  Third, in this small region where the pair
production peaks, spacetime curvature can be neglected.  Fourth, in
the vicinity of the peak, gradient of the electrostatic potential is
constant in magnitude and perpendicular to the zero potential surface,
so the passage of the zero potential surface through the peak itself
guarantees the charge neutrality of the outflow.

The expansion of the pair production rate about the peak is
straightforward and yields,
\ba
w = w_\rmscr{peak} 
\exp \left (-\frac{\pi}{a_0} A_{\mu\nu} \Delta r^\mu \Delta
r^\nu \right )
\label{eq:wrate_exp}
\ea
where $\Delta r^\mu=r^\mu-r_0^\mu$ and $A_{\mu\nu}$ is positive
semidefinite (its nullspace consists of $t-\phi$ plane) and its other
eigenvalues are large compared to the characteristic wavenumbers of the
blackhole, $1/M$ and 1.  The electrostatic injection energy relative to
infinity vanishes at the peak and it also may be expanded near the
peak as $\epsilon = \epsilon_{;\mu} \Delta r^\mu$

Consistent with the approximations mentioned earlier, it is also
immediate to integrate the pair production rate and energy flux over 
space, taking $t=\rmmat{Constant}$ slices,
\ba
\int \sqrt{-g} w d^3 x &=& 4 \pi^2 \sqrt{-g_0} 
 \sigma_1\sigma_2 w_\rmscr{peak} \\
\int \sqrt{-g} |\epsilon| w d^3 x &=& \frac{1}{\sqrt{\pi}} \left |\tilde{A}^{\mu\nu} \epsilon_{;\mu} \right |
 \int \sqrt{-g} w d^3 x 
\label{eq:erate}
\ea
where $\tilde{A}=A^{-1/2}$ and $\sigma_1\sigma_2$ denotes the product of the 
nonzero eigenvalues of $\tilde{A}$.

The energy released (or expended) as the particles travel from where
they form to infinity is given by
\be
\left. E\right |_\infty - \left. E\right |_\rmscr{initial} = 
m \left [ \sqrt {  \frac{\left ( \eta_\mu \xi^\mu \right )^2}{\xi_\mu
\xi^\mu} - \eta_\mu \eta^\mu } - 1 \right ] + \frac{e Q'}{2 M} \left (
\eta_\mu \eta^\mu - 1 \right ) - \frac{e B_0}{2} \eta_mu \xi^\mu. 
\ee
$\epsilon$ denotes the component of the energy proportional to the
charge of the particle.  In Boyer-Lindquist coordinates, this
component is given by
\be
\epsilon = e \left ( \frac{Q'}{M^2} \frac{M^2 r}{\Sigma} + B_0 \frac{2
M^2 r \sin^2 \theta}{\Sigma} \frac{a}{M} \right ).
\ee
$\epsilon$ vanishes on the conical surfaces
\newcommand{\siz}{\sin^2\theta_0}
\be
\siz = -\frac{1}{2} \frac{Q'}{M^2} \frac{M}{a} \frac{1}{B_0} = -\frac{Q'}{\QW}
\ee
The component of the energy proportional to the mass of the particle
insures that within a thin region bounded by
\newcommand{\ciz}{\cos^2\theta_0}
\be
\left | \sin^2\theta - \siz \right | \approx \frac {r^2 + a^2 \ciz}{\QW r} \frac{m}{|e|}.
\ee
particles of neither charge can escape.  Near a stellar-mass black hole,
this region is on the order of an electron Compton wavelength in
thickness; therefore, $m$ may be neglected compared to $e$, leaving
only the electrostatic contribution to the energy for consideration.  The
expansion of $\epsilon$ near the peak yields
\be
\epsilon_{;\mu} = \delta_\mu^\theta \frac {r_0 \sin 2\theta_0}{ r_0^2+a^2 \ciz} e \QW.
\label{eq:epsilon}
\ee

The location and width of the peak is found by numerically
evaluating the pair production rate for a given value of $B_0$ and a
variable value of $Q'/M^2$.   $Q'/M^2$ is varied until the peak
coincides with the zero $\epsilon$ surface.  This procedure works well
for $a \lesssim 0.7 M$.  For larger values of $a$, the peak becomes too
broad and splits in two, so the assumptions above are no longer valid.
However, since the
peak {\em is} broad, the direct numerical technique outlined earlier
gives reliable results.

\begin{figure}
\plottwo{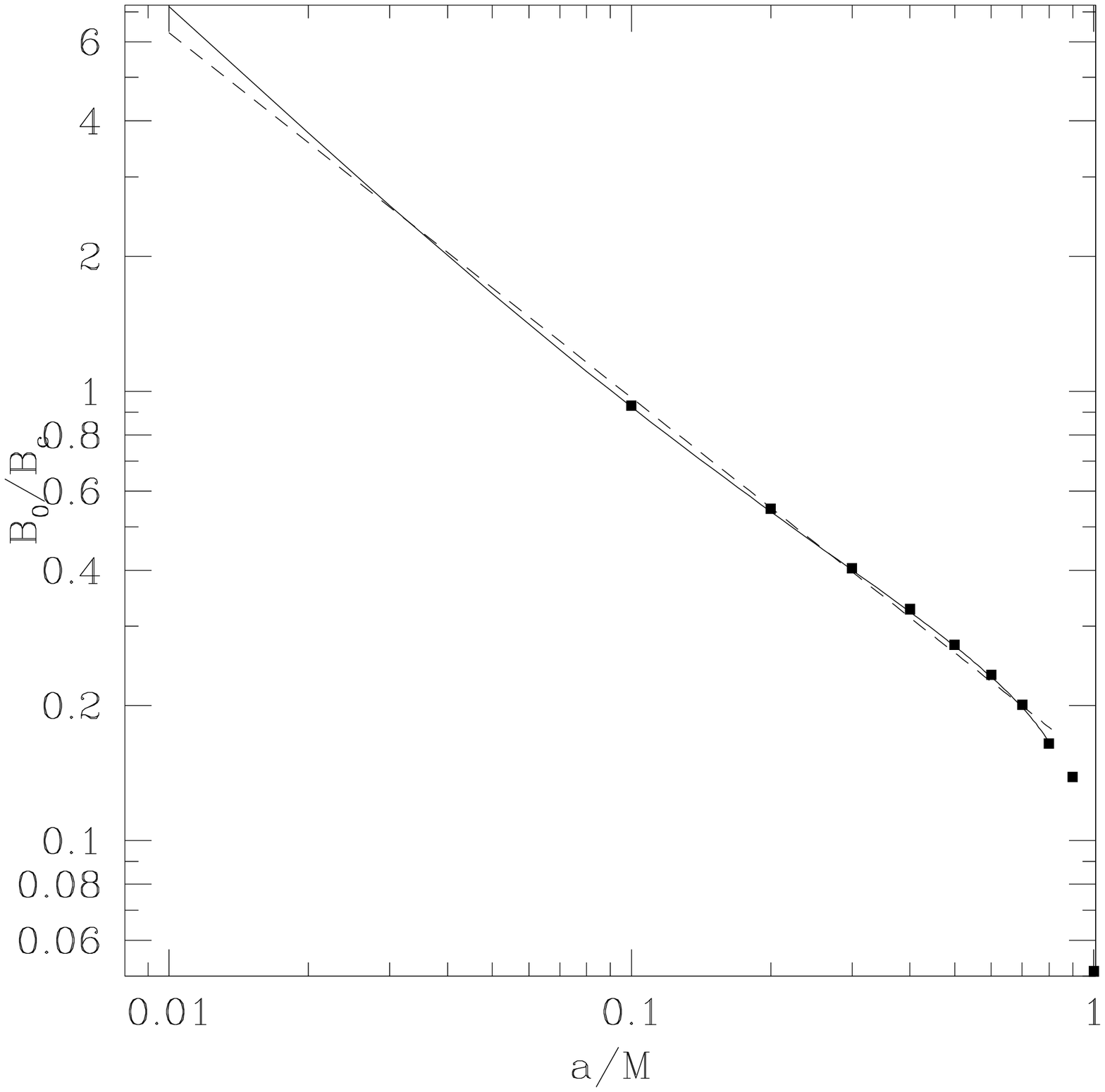}{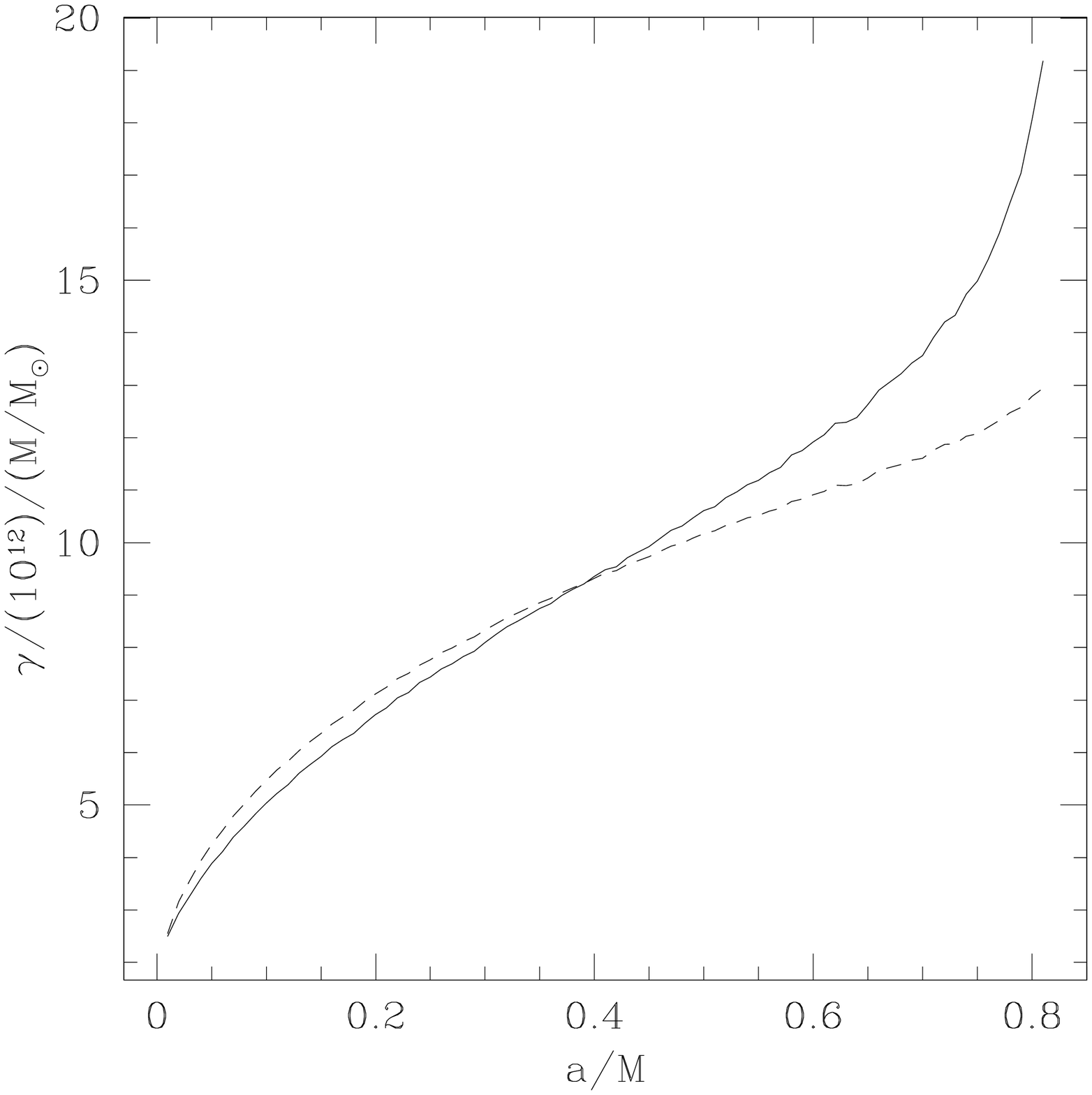}
\caption{The left panel shows the strength of imposed magnetic field
to produce a luminosity of $10^{50} (M/M_\odot)^4$ erg s$^{-1}$ for a
given value of $a/M$.  The solid squares give the numerical results. 
The solid line in right panel traces the mean 
value of $\gamma$ of the primary particles and the dashed line traces the
approximate formula \eqref{gammaapprox}.  }
\label{fig:a_mod1}
\end{figure}

\begin{figure}
\plottwo{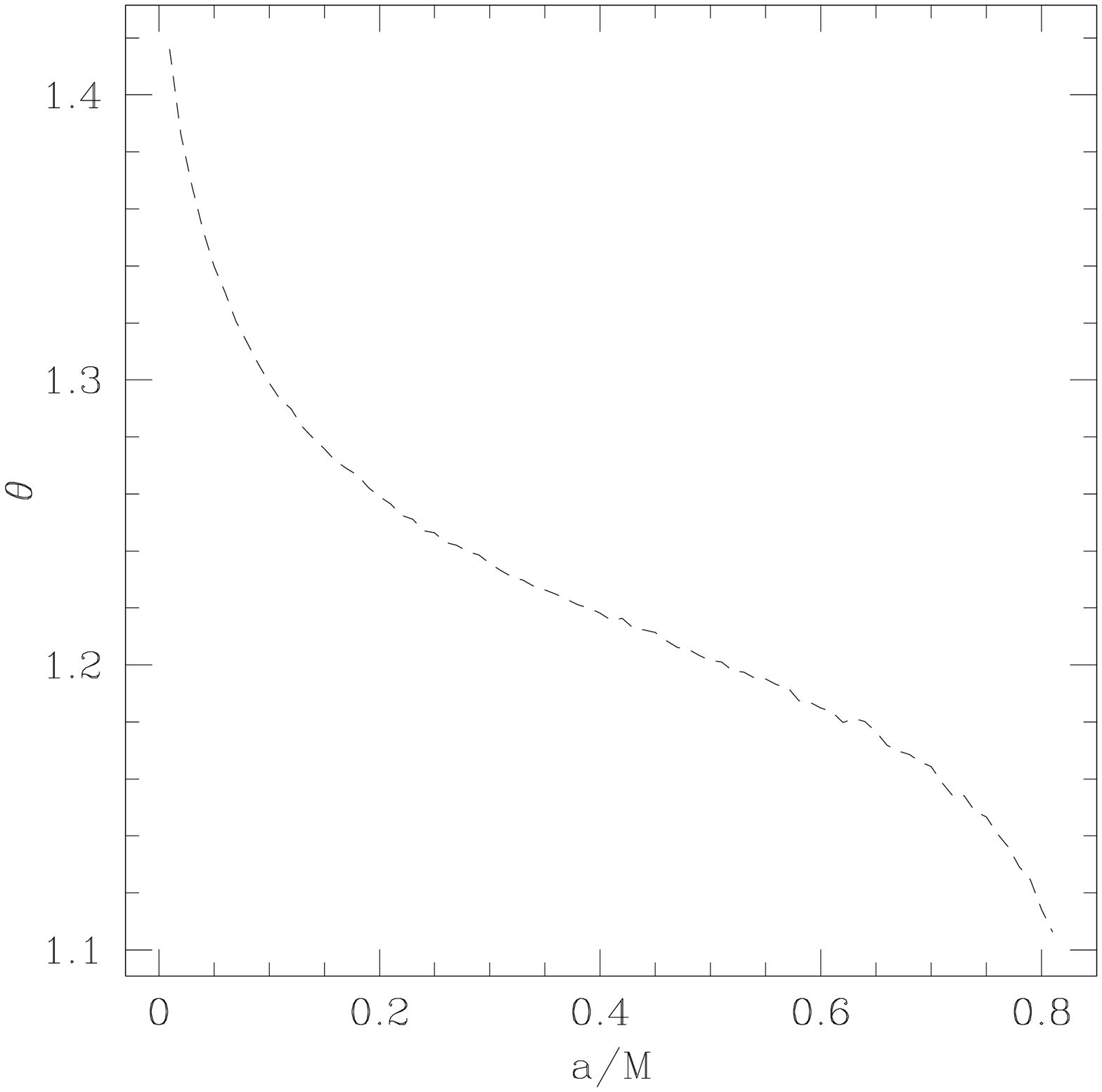}{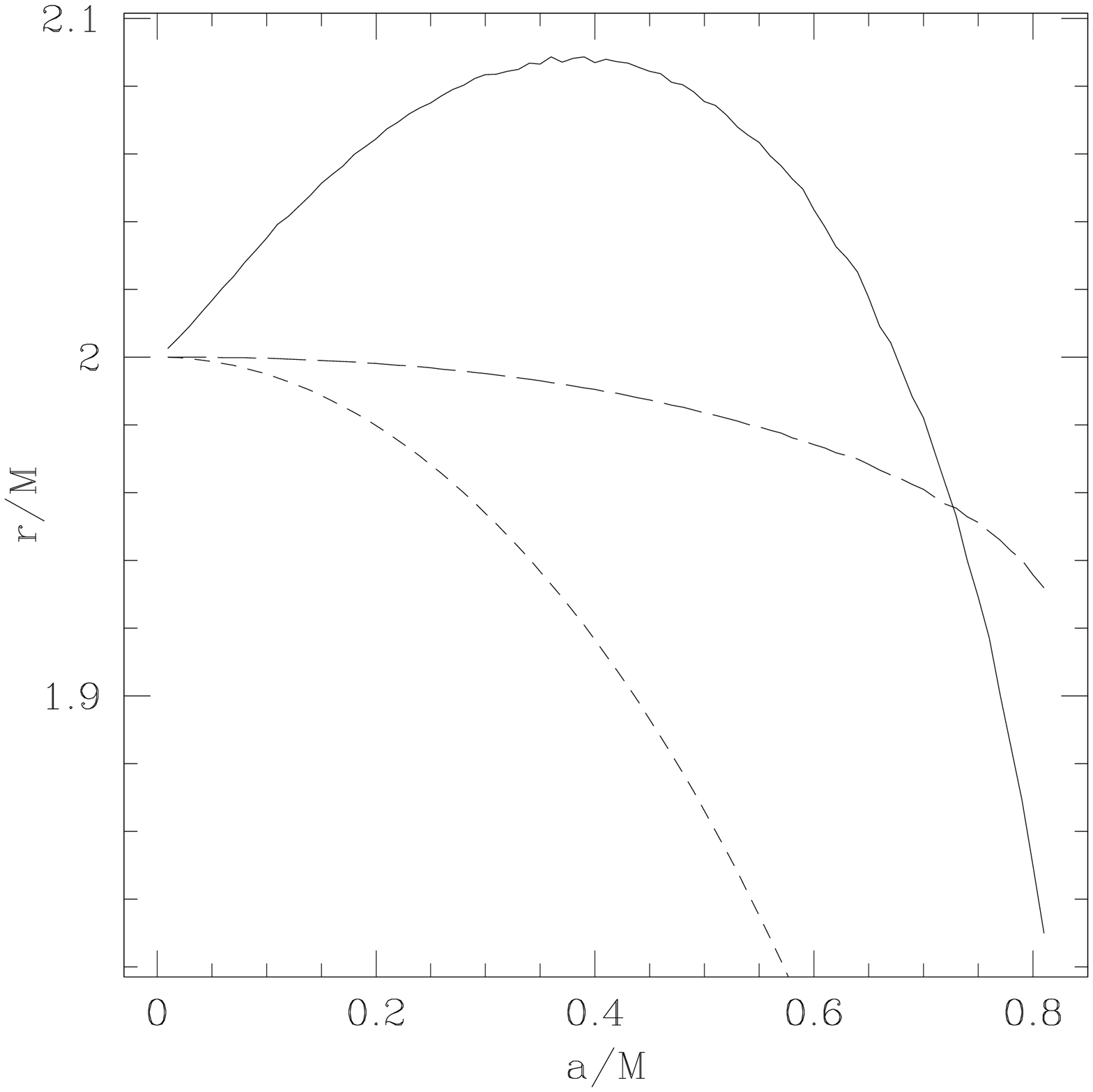}
\caption{The position of the peak of the pair
production as a function of the angular momentum of the black hole.
In the right panel, the radius of the peak is traced by the solid
line, the radius of the horizon by the short-dashed line, and the
radius of the static limit at the latitude of the peak by the
long-dashed line.}
\label{fig:a_mod2}
\end{figure}

\figref{a_mod1} depicts the value of $B_0$ to produce a luminosity of
$10^{50} \left ( M/M_\odot \right )^4$ erg s$^{-1}$ as a function of
$a$ for $0.01 \leq a \leq 0.81$ and the mean value of $\gamma$ for the
primary particles. 
The value of $B_0$ is well fit by a power-law such that 
$B_0 = B_c (10 a/M )^{-4/5}$.  The luminosity increases exponentially with $a$ 
and superexponentially with $B_0$.  A convenient fitting formula is 
\be
\log_{10} \left (\frac{L}{10^{30} \rmmat{erg s}^{-1}} \right ) = 
200 \frac{a}{M} \left ( \frac{B}{B_c} \right )^{5/4} + 4 \log_{10} \left ( \frac{M}{M_\odot} \right ).
\ee
The location of the pair production and typical escape energies of the 
pairs also changes with the angular momentum of the hole.  \figref{a_mod2} 
depicts the colatitude of the peak and the radius of the peak.  A comparison 
of the right panel of \figref{a_mod1} with the left panel \figref{a_mod2} 
verifies that the typical value of $\gamma$ depends rather simply on the 
colatitude of the peak and $a$ through \eqref{erate}.  For small values 
of $a$ ($a \lesssim 0.6 M$), one eigenvector of the matrix
$\tilde{A}$ points in the 
$\theta$-direction, the value of $\gamma$ is approximately given by
\be
\gamma \approx \frac{2\sigma_\theta}{\sqrt{\pi}}
\frac {a r_0 \sin 2\theta_0}{ r_0^2+a^2 \ciz} \frac{B_0}{B_c} \frac{M}{\hbar/mc}
\label{eq:gammaapprox}
\ee 
where $\sigma_\theta \approx 0.055$.  For larger values of $a$, the
eigenvectors rotate away from the $r-$ and $\theta-$directions, 
and the approximation is poorer.  

\begin{figure}
\plotthree{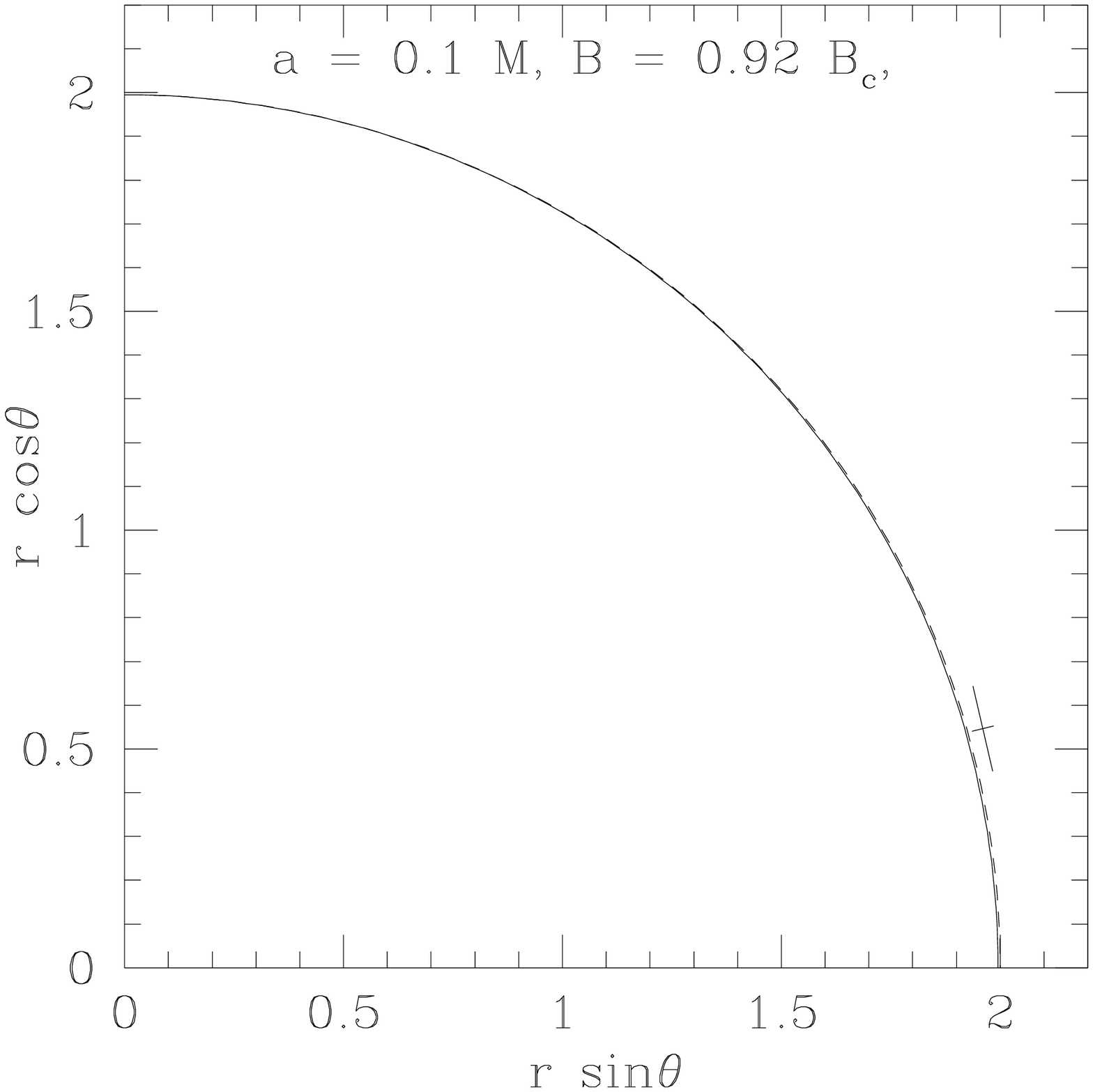}{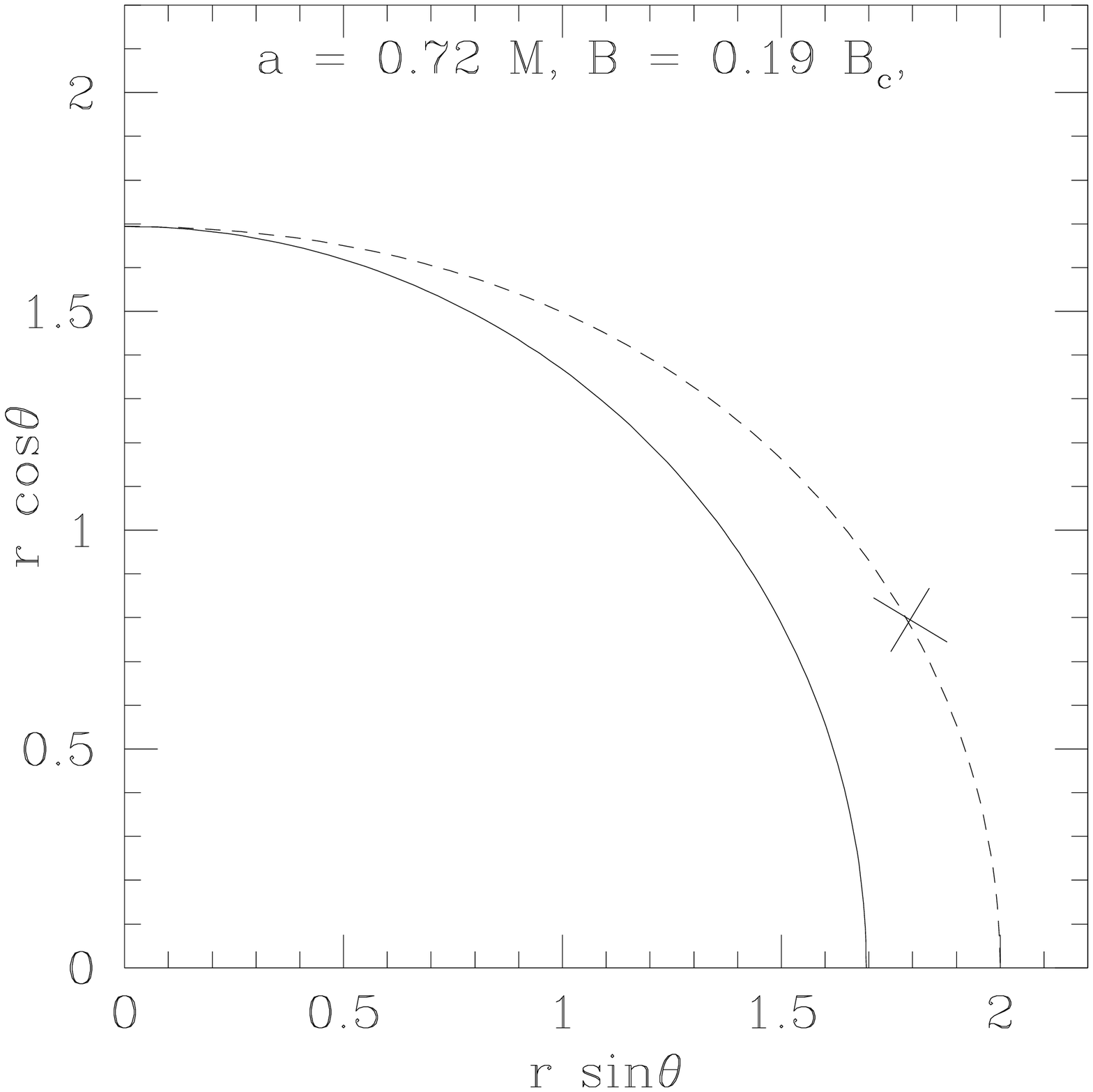}{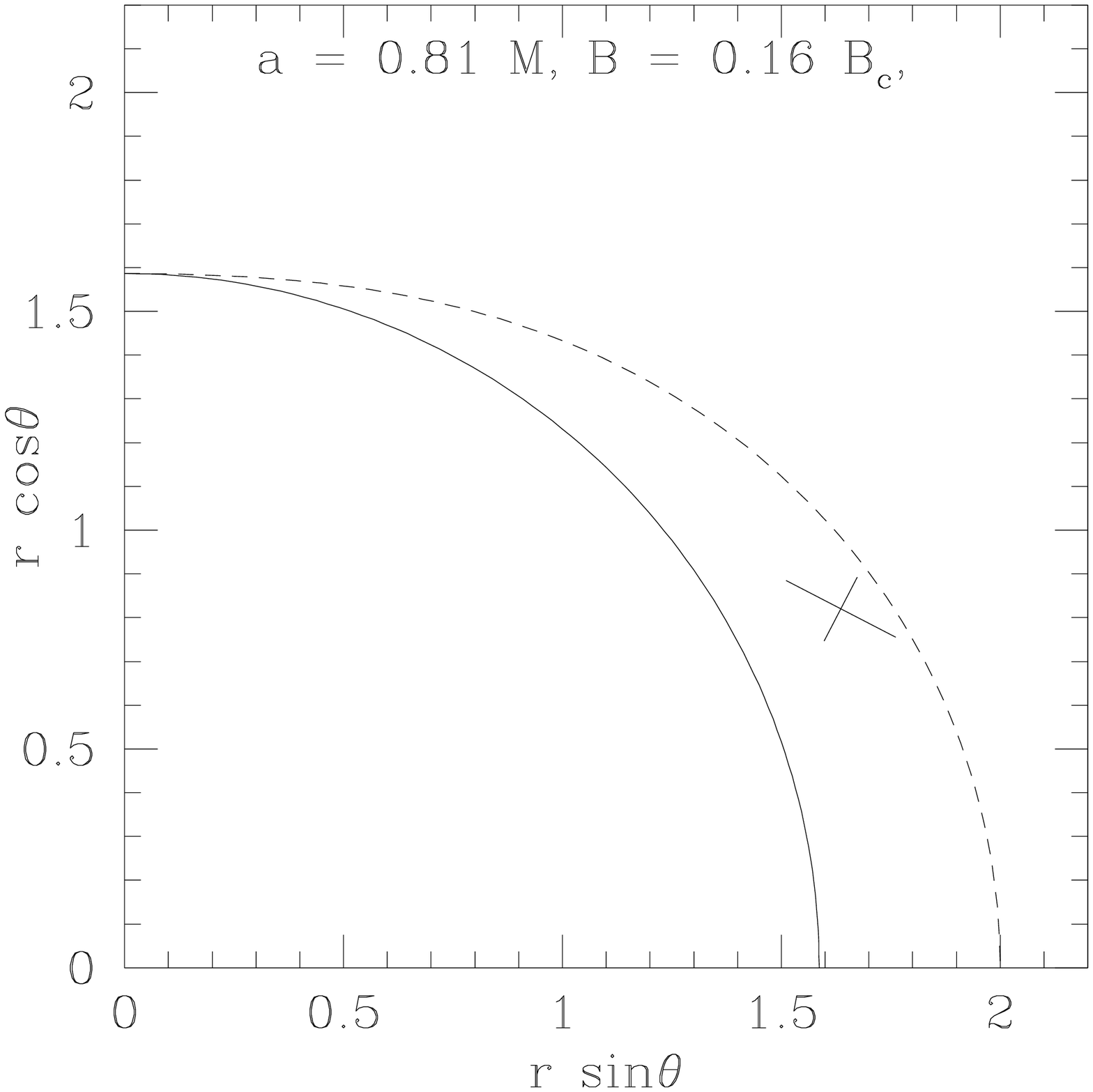}
\caption{The position and principal axes of the peak.  For $a=0.1 M,
0.72 M$ and $0.81 M$.  The peak crosses the static limit at
 $a\approx 0.727 M$.}
\label{fig:peaks}
\end{figure}

The pair production peaks near the equatorial plane for small values
of $a$ and moves toward the poles as $a$ increases.  Meanwhile the
peak remains outside the static limit until $a\approx 0.7 M$ then 
crosses the static limit and moves toward the horizon (\cf \figref{a_mod2}).
\figref{peaks} illustrates how the analytic treatment becomes
unreliable for high values of $a$.  As $a$ approaches $M$, the peak
becomes broad, and the assumptions which support the analytic treatment 
become invalid.

\begin{figure}
\plottwo{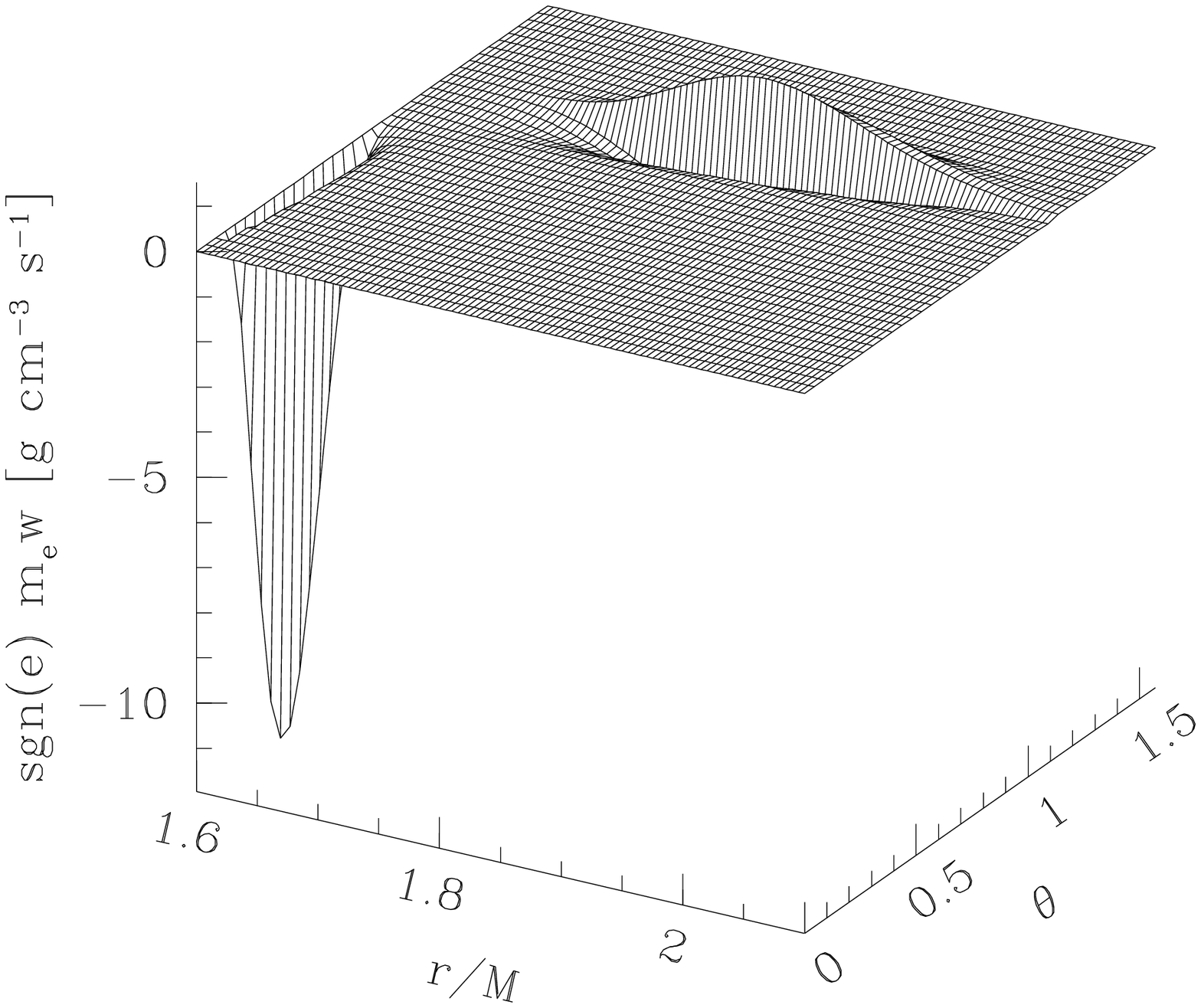}{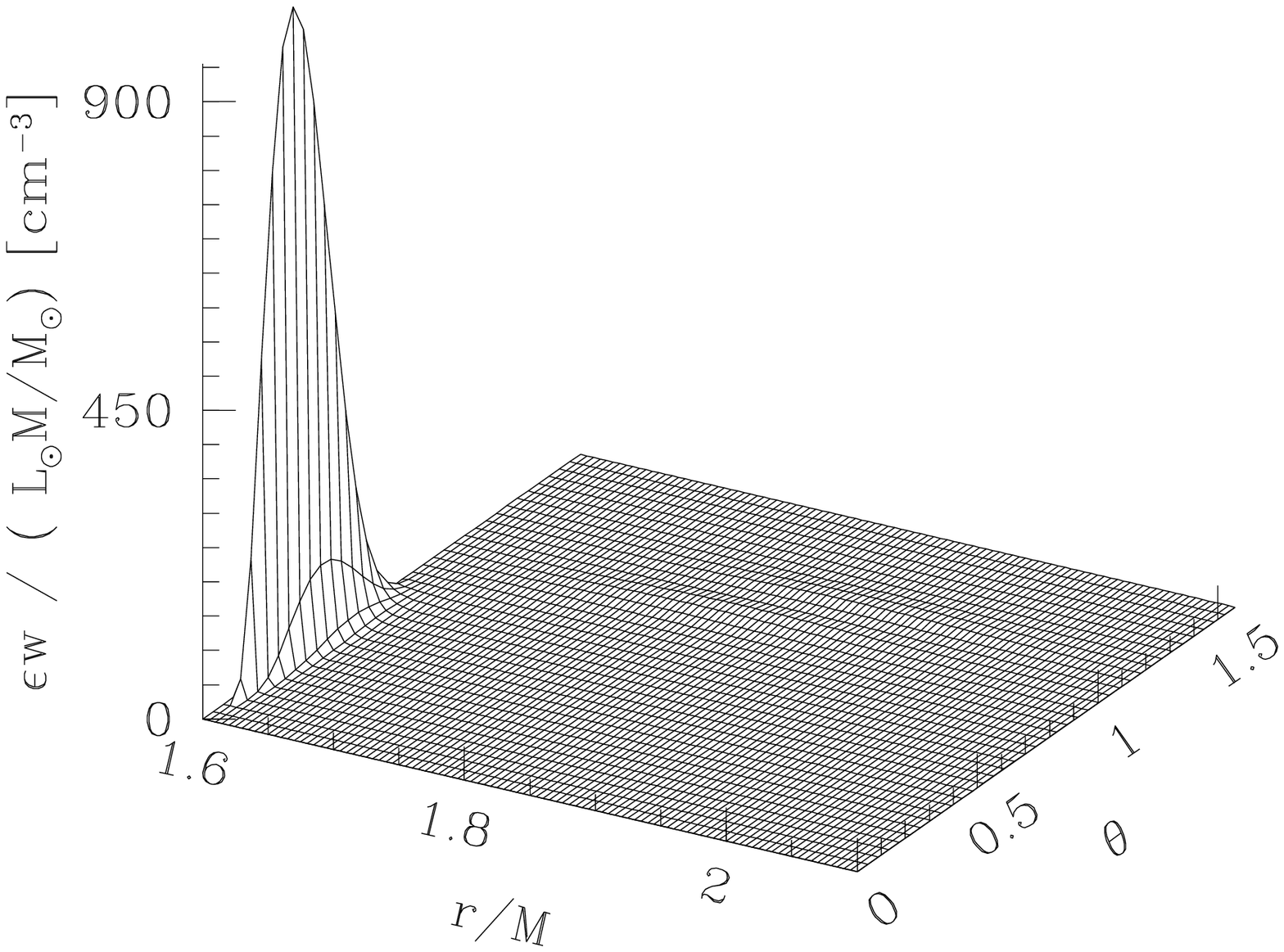}
\caption{The left panel shows the net pair production rate for $a=0.8 M$ 
with $B_0=0.165
B_c$ and $Q'/M^2 = -0.211 B_c$.  The right panel shows the pair-production luminosity.}
\label{fig:nratet0.8}
\end{figure}

For $a>0.735 M$, the value of $I$ on the horizon near the spin axis
becomes negative since $Q'/M^2 < (-4 + 2\sqrt{3}) B_0$; consequently,
for larger values of $a$ pairs are produced both in the main peak and
in the polar regions near the horizon.  This polar component does not
contribute significantly to the total pair production
until $a \sim 0.8 M$  \figref{nratet0.8} depicts the net pair
production rate near a black hole with $a = 0.8 M$.  The main peak is
quite broad with positrons escaping nearer to the equator and
electrons at higher latitudes.  The horizon component of the pair
production consists exclusively of electrons.  Although locally the rate is
quite large, only a small volume is active, so its total contribution
is small.
\section{Conclusions}

The vacuum surrounding a magnetized, rotating black hole is unstable
to pair production if the imposed field approaches $B_k=m^2
c^3/(e\hbar) \approx 4.4 \times 10^{13}$~G $\approx
1.3\times10^{-11}$~cm$^{-1}$.  If the mass of the black hole $M$ is
much less then $10^{11}$~cm or $10^6$~M$_\odot$, the field does not
contribute significantly to spacetime curvature.  For the process as
outlined to operate, the vicinity of the black hole must initially be
free of charge, so the large potential gap and strong electric field
remain stable for subcritical fields; therefore, one would not expect
this simple description to apply astrophysically.  

Adapting the model of \jcite{Goldreich and Julian}{Gold69} to the case
of a rotating black hole provides an estimate of the charge density
necessary to short the electric field.  Two natural definitions of the
angular velocity are that of the zero-angular-momentum frame at the
horizon and at the peak of the pair production.  Both choices result
in the constraint that the electron density must exceed $2-4 \times
10^{15}$ cm$^{-3}$ for a one-solar-mass black hole to short the
electric field.  This is several orders of magnitude larger than the
Goldreich-Julian density typical for rotating neutron stars.  Even if
the initial charge density does exceed this value, the vacuum case
provides important insights.

Van Putten's estimates for the pair production luminosity fall short
of those calculated here by several orders of magnitude
\cite{vanP00a,vanP00b} and depend differently on the mass of the black
hole and the strength of the magnetic field imposed.  Since for weak
fields the luminosity depends superexponentially on the imposed
magnetic field, agreement can be achieved in the gross properties of
the models by slightly varying the magnetic field strength.  For
supercritical fields the pair production rate locally increases as
$B_0^2$, and the electrostatic injection energy, $\epsilon$, is
proportional to $B_0 M$; consequently, the total luminosity increases
as $B_0^3 M^4$ -- van Putten argues that the total luminosity
increases as $B_0^2 M^2$.  Reconciling these differences is difficult
as this report and van Putten's work treat the underlying physical
processes differently.

Some natural extensions to this work are a treatment of the back
reaction of the outflow on the spin of the black hole and the source
of the external magnetic field.  By including a more realistic
description of the magnetic field far from the hole which would likely
include a model for its source, the beaming of the relativistic jet
could be determined.  These developments would constrain the duration
of the emission, the nature of its onset and its possible modulation.
The pairs will naturally produce secondary particles as they travel
along the curved magnetic field lines.  These secondaries, the
primaries or previously present material could form a magnetosphere
around the black hole, causing the electromagnetic field to evolve
toward a force-free configuration (\eg \cite{LeevanP00}).

Rotating black holes coupled to strong magnetic fields naturally
produce a highly relativisitic, columnated outflow with a total
luminosity which can easily exceed $10^{50}$ erg/s.  The electrons and
positrons are initially separated.  For $a \gtrsim 0.8 M$, if the
applied field is parallel (antiparallel) to the angular momentum of
the black hole, electrons (positrons) will escape from polar regions
near the horizon from a region near the equatorial plane within the
static limit.  Positrons (electrons) escape only from the equatorial
region.  For small values of $a$, both electrons and positrons escape
from a small region outside the static limit whose latitude depends on
the value of $a$.

\acknowledgments
Support for this work was provided by a Lee A. DuBridge postdoctoral 
fellowship and the National Aeronautics and Space Administration
through Chandra Postdoctoral Fellowship Award Number PF0-10015 issued
by the Chandra X-ray Observatory Center, which is operated by the
Smithsonian Astrophysical Observatory for and on behalf of NASA under
contract NAS8-39073.  I would like to thank M. van Putten, L. Hernquist, 
and R. Narayan for useful discussions.

\bibliography{ns,qed,leads,physics,mine,gr} 
\bibliographystyle{prsty}

\end{document}